# Energy-Efficient Routing Protocol Based on Multi-Threshold Segmentation in Wireless Sensors Networks for Precision Agriculture

Yin-Di Yao, Xiong Li, *Student Member, IEEE*, Yan-Peng Cui, *Student Member, IEEE*, Jia-Jun Wang, and Chen Wang

*Abstract*—Wireless sensor networks (WSNs), one of the fundamental technologies of the Internet of Things (IoT), can provide sensing and communication services efficiently for IoT-based applications, especially energy-limited applications. Clustering routing protocol plays an important role in reducing energy consumption and prolonging network lifetime. The cluster formation and cluster head selection are the key to improving the performance of the clustering routing protocol. An energy-efficient routing protocol based on multi-threshold segmentation (EERPMS) was proposed in this paper to improve the rationality of the cluster formation and cluster heads selection. In the stage of cluster formation, inspired by multi-threshold image segmentation, an innovative node clustering algorithm was developed. In the stage of cluster heads selection, aiming at minimizing the network energy consumption, a calculation theory of the optimal number and location of cluster heads was established. Furthermore, a novel cluster head selection algorithm was constructed based on the residual energy and optimal location of cluster heads. Simulation results show that EERPMS can improve the distribution uniformity of cluster heads, prolong the network lifetime and save up to 64.50%, 58.60% and 56.15% network energy as compared to RLEACH, CRPFCM and FIGWO protocols respectively.

*Index Terms*—Clustering routing protocol, cluster head selection, multi-threshold segmentation, network energy consumption, wireless sensor networks (WSNs).

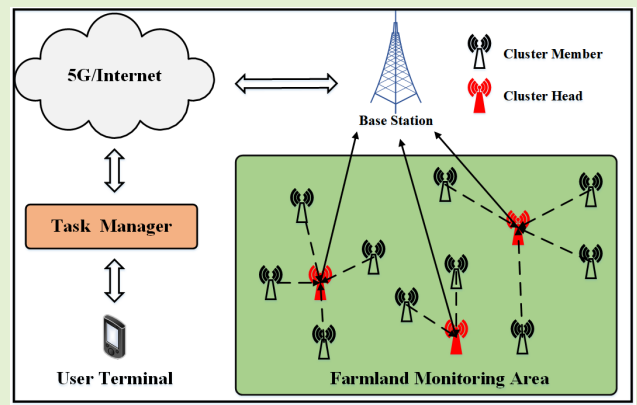

## I. Introduction

THE rise of IoT has greatly promoted the intelligence of transportation, home, precision agriculture and healthcare systems [1]. In particular, IoT-based precision agriculture [2] has aroused extensive research in recent years. Fig. 1 shows a precision agricultural irrigation system [3] where a large number of power-limited wireless sensor nodes with communication and computing capabilities are randomly deployed in a circular area to collect and process parameters like air temperature, soil moisture, PH, etc., and then forward it to the base station (BS) for making smarter decision so that the irrigation can take place automatically. In such an application, it is difficult to replace the battery of sensor nodes and realize maintenance of the network [4]. Once a node fails due to exhaustion of node energy, resulting in the loss of complete farmland monitoring data, it will bring irreparable losses to the network [5]. So it is of great significance to reduce node energy consumption mainly from the long-distance data transmission and inevitable data acquisition. Therefore, how to design an energy-efficient routing protocol to reduce the transmission energy consumption of nodes and prolong the network lifetime has been a crucial question.

Clustering routing protocol [6] has emerged as the most energy efficient protocol that divides the sensor nodes into clusters. In clustering routing protocol (see Fig. 1), all nodes are divided into multiple clusters, and a cluster head (CH) is selected in the cluster to fuse and forward data from

Manuscript received January 15, 2022; revised February 3, 2022; accepted February 8, 2022. Date of publication February 10, 2022; date of current version March 31, 2022. This work was supported in part by the National Natural Science Foundation of China under Grant U1965102, in part by the Science and Technology Innovation Team for Talent Promotion Plan of Shaanxi Province under Grant 2019TD-028, in part by the Communication Soft Science Project (2021-R-47), in part by the Shaanxi Science and Technology Plan Project (2021NY-180), in part by the Key Research and Development Project of Shaanxi Province (2020NY-161), and in part by the Shaanxi University Students' Innovation and Entrepreneurship Project (S202111664121). The associate editor coordinating the review of this article and approving it for publication was Prof. Reza Malekian. *(Corresponding author: Xiong Li.)*

Yin-Di Yao, Xiong Li, Jia-Jun Wang, and Chen Wang are with the School of Communication and Information Engineering and the Shaanxi Key Laboratory of Information Communication Network and Security, Xi'an University of Posts and Telecommunications, Xi'an 710121, China (e-mail: yaoyindi@xupt.edu.cn; leexiong123@163.com; wjjnb0414@163.com; 992566842@qq.com).

Yan-Peng Cui is with the School of Information and Communication Engineering, Beijing University of Posts and Telecommunications, Beijing 100876, China (e-mail: cuiyanpeng94@bupt.edu.cn).

Digital Object Identifier 10.1109/JSEN.2022.3150770

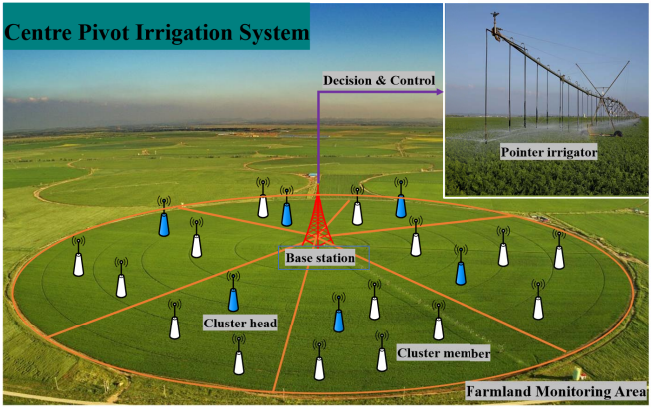

Fig. 1. IoT-based precision agriculture system using WSNs.

other cluster members (CMs), and then conveying it to BS. By grouping the nodes into clusters with the assistance of data fusion mechanism [7], energy efficiency is improved since the data transmission volume sent to BS is considerably reduced and the communication distance between nodes (CH to CMs or CHs to BS) is optimized.

However, predominantly, the CH undertakes a huge number of data forwarding and fusing tasks and runs out of energy earlier than cluster members (CMs). Furthermore, if the number of CMs served (managed) by each CH is very different, the imbalance of energy consumption among cluster heads will further accelerate the death of some CHs. Additionally, network energy consumption is closely related to the location and number of cluster heads. When the number of CHs is large, the clustering efficiency and utilization rate of CHs are low, which makes it difficult to minimize energy consumption. Conversely, each CH undertakes a large amount of data fusion and long-distance forwarding tasks, and the communication overhead will surge. In addition, if the distance between the CH and BS is close (or far), the communication energy consumption between the CMs and CH (or CH and BS) will be larger. Worse, to our best knowledge, the extensive research [8] [9] [16] often considers the optimal number of CHs and ignores the influence of the difference of network size on its optimal solution, and even less discusses the optimal value of the distance between CHs and BS. Furthermore, in order to enhance the balance of CH energy consumption, clusters are usually divided into different sizes (the cluster size near BS is small, and vice versa) to avoid CHs near BS taking on additional data forwarding tasks resulting in accelerating their death [10], [11]. But it failed to solve the problem of unbalanced energy consumption caused by the huge difference in the number of CH loads. Therefore, how to optimally set the number of cluster heads, optimize the distribution of cluster heads in the monitoring area and ensure cluster head load balance so as to maximize the network lifetime and make the network energy efficient are important challenges faced by clustering routing protocols.

In this paper, we use the multi-threshold segmentation algorithm (specifically, Otsu algorithm [12]) to develop and analyse an innovative energy-efficient routing protocol based on multi-threshold segmentation (EERPMS). Sensor nodes are grouped into clusters according to the principle of the node angle and number variance between clusters. The results of node clustering based on Otsu algorithm can effectively improve the distribution uniformity and the load balance of CHs. In addition, a CH selection algorithm based on the optimal location of CHs and the nodes residual energy is proposed to optimize the selection of CHs so as to reduce network energy consumption and extend network lifetime.

Briefly, the main contributions of our research are summarized as follows.

1) By constructing the relationship between multi-threshold image segmentation and node clustering, the node clustering problem is transformed into the problem of solving the optimal segmentation threshold. Combining the Otsu algorithm with node angle and number, a cluster forming method based on the variance of node angle and number between clusters is proposed. It can improve distribution uniformity and load balance of CHs.
2) The communication models between nodes in different scenarios are discussed, and finally the scenarios satisfying the free space model between nodes are determined to avoid a large amount of energy consumption between nodes aroused by the multipath fading model.
3) The closed-form expressions of the optimal location and number of CHs are derived by revealing the mapping relationship among network size, the location and number of CHs, and network energy consumption. Based on the closed-form expressions, we set the objective function with the node residual energy and the optimal CH location to optimize the selection of CHs.
4) Combined with the node clustering method based on multi-threshold segmentation and the CH selection method based on the optimal CH location, an energy-efficient routing protocol based on multi-threshold segmentation is proposed to reduce node energy consumption and prolong network lifetime.

The rest of the paper is structured as follows: Section II reviews the previous related work. The assumptions of system model and energy consumption model are introduced in Section III. Section IV describes and analyzes the proposed work. In Section V, the simulation results and the causes of performance differences are analyzed in detail. At last, we concluded our researches and discussed the future work.

## II. RELATED WORK

Wireless sensor technologies are increasing interest in the precision agriculture domain. The major challenges of WSNs are developing low-cost and energy-efficient routing protocols with simple operation and high reliability. The energy-efficient WSNs routing protocols for precision agriculture have been presented as various proposed approaches like region-based static clustering approach and dynamic clustering approach, etc [4]. However, due to space restriction, only some of these proposed approaches are presented in this section.

The static clustering approach where the network area is first divided into several fixed sub-regions, and deployment of different types of nodes in the respective regions according to their energy level provides efficient utilization of the total coverage area. Then, some CHs are selected for those outer sub-regions using fuzzy logic technique or swarm intelligent optimization method, etc. Sonam Maurya *et al.* [13] developed a Region-Based Hybrid Routing Protocol (RBHR) where the monitoring area is divided into five sub-regions and two types of nodes are deployed. Low energy nodes in region 1 send their data to BS directly and high energy nodes in regions 2-4 transmit their data by using transmission of CH selected by fuzzy logic. Parameters for CH election like the node distance to BS and residual energy at each node are considered. To some extent, it can improve balancing of network energy thanks to the introduction of heterogeneous nodes and hybrid transmission modes. The similar method of region division and CH selection was used in Threshold Sensitive Region-Based Hybrid Routing Protocol (TS-RBHR) [14]. The data transmission of TS-RBHR occurs only when the temperature or soil moisture level of field reaches the desired threshold, which can reduce the data transmission volume so as to save network energy. However, such static clustering approach achieves the effect of balancing energy consumption by artificially deploying heterogeneous nodes in different regions, but the flexibility and scalability are very poor. In [15], BS is supposed to movably collect data from every CHs in sub-grids and the objective function is set to optimize end-to-end packet loss rate. This can efficiently minimise the energy consumption of CHs, but that data receiving mode of mobile base station is not suitable for all agricultural scenarios, especially those under harsh natural conditions.

As a typical dynamic clustering approach, Low Energy Adaptive Clustering Hierarchy (LEACH) [16] dynamicly elects CHs according to factors such as times that a node is elected as CH, probability threshold and current network rounds, etc. Its advantage lies in that each node has the same opportunity to act as a CH, thus avoiding premature death caused by a certain node undertaking the task of forwarding data for many times. Moreover, the relationship between network energy consumption and CH number is constructed at first. By redefining the probability threshold based on the node residual energy and the optimal number of CHs, the residual based LEACH routing (RLEACH) was proposed in [17] where the chance of low energy nodes being CHs is reduced resulting in the extension of network lifetime. Based on the three-layer hierarchy structure, Lee [18] *et al.* designed a hybrid hierarchical clustering approach where a novel grid head is designed to transmit the data from layer-1 CHs.

Thanks to the support of AI technology, metaheuristic algorithms, fuzzy clustering and fuzzy logic technologies have been applied to more efficient dynamic clustering routing protocols. In [19], by improving the weights of the top three wolves based on fitness values by setting fitness functions about residual energy and distance from the BS, fitness-value based improved grey wolf optimizer (FIGWO) was developed, in which CHs with more high energy and near BS is selected for shortening the communication distance between CHs and BS. Panchal *et al.* [20] developed a FCM clustering protocol where the sensor nodes are grouped into clusters with a degree of belonging to each cluster rather than hard partitioning them into only one cluster. However, FCM-based clustering protocol can optimize the distance between intra-cluster nodes, but not between CHs and BS. Hence, CHs energy consumption is not saved. Reference [7] proposed the outstanding strategy of enhanced cluster head selection to minimizing data redundancy where the redundant data is significantly reduced owing to the sleeping-waking mechanism, and the data communication distance is optimized thank to novel manner of CH selection (fuzzy IF-THEN rules related with energy and local distance). However, although the above mentioned methods are excellent in some performances, the energy consumption balance is not considered.

Consequently, the energy consumption balance has attracted extensive research. In particular, the energy consumption balance of circular region where the BS are positioned at center of the circular region divided into circular sectors. Neeti *et al.* [21] proposed an adaptive sectoring and cluster head selection based multi-Hop routing algorithm in which CH was selected once in a round and the frequency depends upon its distance from BS so that levels near BS receive more data. Arghavani *et al.* [22] studied an optimal multi-hop communications for a circular area surrounding a sink in which the network was partitioned into nearly the same size clusters and optimal number of clusters (namely, CH) was discussed under multi-hop routing. An Energy-efficient and Coverage-guaranteed Unequal-sized Clustering (ECUC) scheme for the circular area was proposed in [23], its innovation lies in that the energy consumption after clustering was reduced by considering the node power consumption before clustering using the cluster size optimization. In [24], the cluster formation method is that the larger numbers of clusters were formed closer to BS and reduce as one moves away from the BS. Furthermore, the optimal CHs were selected based on the accumulative weights related to the residual battery, node degree and centrality. But the selection of relay point was not studied so that the overall network lifetime cannot be effective extension.

Table I summarizes the existing clustering routing protocols with important features and compared with the proposed protocol.

## III. Preliminaries
### A. System Model and Assumptions

$N$ sensor nodes are randomly deployed in a circular agricultural monitoring area of radius $R$ similar to [22], [25]. Some reasonable assumptions have been made as follows.

1) All nodes are randomly deployed in the monitoring area and node distribution function follows uniform distribution.
2) All nodes and BS are fixed after deployment.
3) Each node has the same initial energy and communication range, and can obtain their geographical position and angle relative to BS by GPS or other positioning algorithms.
4) All nodes with limited power have the ability to calculate, process and forward data.

TABLE I
SUMMARY OF THE RELATED WORK

| Protocol | Methodology and Difference |
|---|---|
| RBHR [13] | Region-based routing and fuzzy logic-based CH selection |
| TS-RBHR [14] | Region-based routing and threshold-sensitive data transmission mode |
| MS-routing-$G_i$[15] | Mobile BS and random CH selection method |
| LEACH [16] | Dynamic clustering and probability-based CH selection |
| RLEACH [17] | Dynamic clustering and Probability-based CH selection with node residual energy |
| HHCA[18] | Three-level dynamic clustering and FCM-based cluster formation |
| FIGWO [19] | Metaheuristic algorithms-based and CH selection with node energy and CH location |
| FCM [20] | FCM-based cluster formation |
| ECH [7] | random CH selection and optimal CH number |
| Neeti Jain [21] | Multi-Hop routing algorithm and unequal-Clustering |
| Mahdi Arghavani [22] | Optimal number of clusters under Multi-hop routing |
| ECUC [23] | Unequal-clustering and cluster size optimization for reducing energy consumption |
| IUCR [24] | Unequal-clustering and CH selection with residual energy, node degree and centrality |

5) BS is located in the center of the monitoring area and it can get the information of all nodes.

All the above assumptions are reasonable and feasible. In the agricultural monitoring scene, nodes are fixed in monitoring points to monitor soil moisture data. BS is usually arranged in the regional center to process and analyze the data and then control the mobile pointer center irrigator to complete the irrigation task.

### B. Energy Consumption Model

The energy consumption used by us follows the popular model given in [16] to calculate energy consumption of sensor nodes. The energy consumed in the transmission of $l$-bit data at the distance $d$, is represented by $E_{Tx}$ and it is calculated in (1). If $d \leq d_{th}$, the energy consumption model is called free space model ($d^2$ energy dissipation). Conversely, it is the multipath fading model ($d^4$ energy dissipation). The distance threshold $d_{th}$ is computed by (2).

$$E_{Tx}(l,d) = \begin{cases} lE_{elec} + lE_{fs}d^2, & if \quad d \leq d_{th} \\ lE_{elec} + lE_{mp}d^4, & if \quad d > d_{th} \end{cases} \quad (1)$$

$$d_{th} = \sqrt{E_{fs}/E_{mp}} \quad (2)$$

where, $E_{elec}$ denotes the coefficient of the transmitter or receiver. $E_{fs}$ and $E_{mp}$ represent the energy consumption coefficients of free space model and multipath fading model, respectively.

Then, the energy consumption $E_{Rx}$ of $l$-bit data received by the data receiving module can be calculated by (3).

$$E_{Rx}(l) = lE_{elec} \quad (3)$$

### C. Multi-Threshold Otsu Algorithm

Given that the multi-threshold Otsu algorithm [26] is the basis of cluster formation in EERPMS, we will briefly introduce the principle of the multi-threshold Otsu algorithm now. As one of the most well-known segmentation techniques, it is used to find the optimal threshold of an image to divide the image into many classes according to the maximization of variance between classes.

Firstly, for an gray image with $L$ intensity levels, the probability distribution of the histogram is calculated as in (4).

$$p_i = h_i/N_h, \sum_{i=1}^{N_h} p_i = 1 \quad (4)$$

where, $i_l$ is an intensity level specified in the range of $0 \leq i_l \leq L-1$. $N_h$ is the total number of image pixels. $h_i$ denotes the number of the appearance of intensity $i_l$ in the image indicated by the histogram.

Based on the probability distribution or thresholding value $TH = \{th_1, th_2, \ldots, th_{K-1}\}$, the classes $C$ are determined for $K$-level segmentation, and the class $C_1$ is denoted as $C_1 = \{p_1/\varpi_1, \ldots, p_1/\varpi_{th_1}\}$, other classes can be analogized. $\varpi_k, k = 1, 2, \ldots, K$ are cumulative probability distributions, as it is depicted in (5).

$$\varpi_1 = \sum_{i=1}^{th_1} p_i, \varpi_2 = \sum_{i=th+1}^{th_2} p_i, \ldots, \varpi_K = \sum_{i=th_{K-1}}^{L-1} p_i \quad (5)$$

Then, it is mandatory to find the average intensity levels $\mu_k, k = 1, 2, \ldots, K$ using (6).

$$\mu_1 = \sum_{i=1}^{th_1} ip_i/\varpi_1, \mu_2 = \sum_{i=th_1+1}^{th_2} ip_i/\varpi_2, \ldots,$$

$$\mu_K = \sum_{i=t_{K-1}}^{L} ip_i/\varpi_K \quad (6)$$

Lastly, the Otsu based between-class $\sigma_B^2$ defined by (7).

$$\sigma_B^2 = \sum_{k=1}^{K} \varpi_k(\mu_k - \mu_T)^2 \quad (7)$$

where $\mu_T$ is the total gray mean of the image, as shown in (8).

$$\mu_T = \sum_{k=1}^{K} \varpi_k \mu_k \quad (8)$$

Eq. (9) presents the objective function. Therefore, the optimization problem is reduced to find the intensity level that maximizes (9).

$$F_{Otsu} = Max(\sigma_B^2(TH)), 0 \leq th_i \leq L-1,$$
$$i = [1, 2, \ldots, K-1] \quad (9)$$

### D. Scenario Analysis

In order to reduce the huge energy consumption caused by long-distance communication between nodes, combined with specific precision agricultural irrigation scenarios, the following scenario analysis is proposed to determine the scenario that nodes follow the free space model under the condition of CH relay. *Theorem 1* is presented as follows.

*Theorem 1:* For the circular monitoring area centered on BS, if there exist $K$ cluster heads that make the communication distance between nodes conform to the free space model,

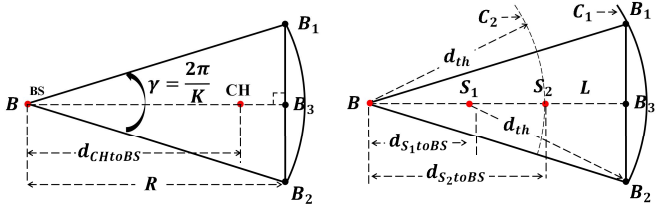

Fig. 2. The shape of a cluster area.

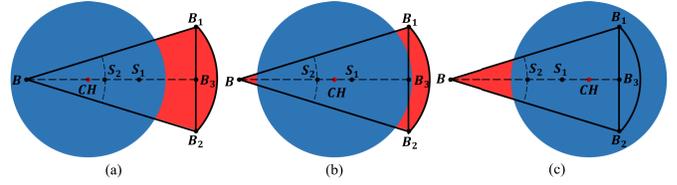

Fig. 3. The location of CH in cluster area when $d_{S_2toBS} < d_{S_1toBS}$.

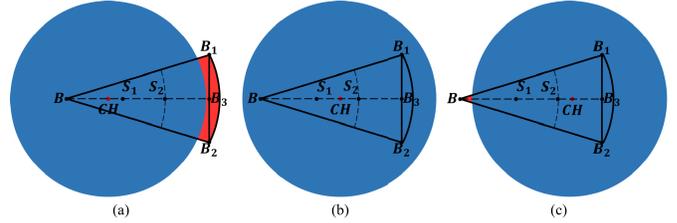

Fig. 4. The location of CH in cluster area when $d_{S_2toBS} \geq d_{S_1toBS}$.

the radius $R$ of the monitoring area should meet the following conditions, as shown in (10).

$$R \leq 2d_{th}cos(\pi/K) \quad (10)$$

*Proof:* For circular monitoring area of radius $R$ with randomly deploying $N$ sensor nodes and $K$ clusters area, if BS is located in the center of the area, assuming that each cluster has the same number of nodes (contains one CH and the N/K-1 CMs), the shape of the clusters are sector (bottom corresponding vertex for BS, apex angle for $\gamma = 2\pi/K$), and CH is located on the bottom mid-perpendicular $L$. The three vertices of the cluster area are denoted as $B$, $B_1$ and $B_2$ respectively, as shown in Fig. 2, and the intersection point between $L$ and edge $B_1B_2$ is denoted as $B_3$.

According to the relationship between the distance between nodes and BS and $d_{th}$, the monitoring area can be divided into the following two types of scenarios:

1) The distance between all nodes and BS shall not exceed $d_{th}$. The scenario is equivalent to that the farthest distance $R$ between all nodes and BS are not exceed $d_{th}$, namely the area radius $R \leq d_{th}$. At this point, the communication model between nodes conforms to the free space model.

2) There are nodes with the distance from BS greater than $d_{th}$ (i.e., $R > d_{th}$). We define that the point on $L$ with distance $d_{th}$ from points $B_2$ is denoted as $S_1$, and the point with distance $d_{th}$ from BS is denoted as $S_2$, $C_1$ and $C_2$ are arcs with radius $d_{th}$ as center of $S_1$ and $S_2$ respectively. The distance between BS and $S_1$ and $S_2$ is denoted as $d_{S_1toBS}$ and $d_{S_2toBS}$ respectively, as shown in (11) and (12). When $d_{S_1toBS} = d_{S_2toBS}$, the radius threshold $R_{th}$ is shown in (13).

$$d_{S_1toBS} = Rcos(\pi/K) - \sqrt{d_{th}^2 - R^2sin(\pi/K)^2} \quad (11)$$

$$d_{S_2toBS} = d_{th} \quad (12)$$

$$R_{th} = 2d_{th}cos(\pi/K) \quad (13)$$

Note that according to the positional relationship between $S_1$ and $S_2$, this scenario can be also divided into the following two cases:

a) $d_{S_2toBS} < d_{S_1toBS}$ (i.e., $R > R_{th}$). As shown in Fig. 3, the red area represents the non overlapping region of blue circle and sector. In this case, regardless of CH is located in where, the distance between the nodes in red shade area and CHs (or BS) is beyond $d_{th}$, and there must exist nodes to communicate with multipath fading model (which may be CH or CMs).

b) $d_{S_2toBS} \geq d_{S_1toBS}$ (i.e., $R \leq R_{th}$). If $d_{CHtoBS} < d_{S_1toBS}$ or $d_{CHtoBS} > d_{S_2toBS}$, that is, when the CH is on the left side of $S_1$ or the right side of $S_2$, as shown in Fig. 4 (a) and (c), there exist nodes located in red area whose communication distance from CH is beyond $d_{th}$. If $d_{S_1toBS} \leq d_{CHtoBS} \leq d_{S_2toBS}$, that is, when the CH is between $S_1$ and $S_2$, as shown in Fig. 4 (b), the communication distance between all nodes does not exceed $d_{th}$, that is, the energy consumption between all nodes conforms to free space model. It is easy to know that the distance between CH and BS should meet the conditions described in (14);

$$d_{CHtoBS} \in [Rcos(\pi/K) - \sqrt{d_{th}^2 - R^2sin(\pi/K)^2}, d_{th}] \quad (14)$$

To sum up, when $R \leq R_{th}$, the communication energy consumption between all nodes is free space model caused by the existence of CH. The proof is completed.

## IV. PROPOSED WORK

In this section, we present the centralized EERPMS protocol, which includes the following phases: clustering formation, CH selection and data transmission. A complete round consists of the above three phases and the operation of our protocol is partitioned into rounds. Fig. 5 shows the overall operation of EERPMS. Note that the first two phases of EERPMS are conducted by BS, and then BS informs all nodes to complete the network construction among nodes. Especially, when the number of dead nodes has changed, EERPMS protocol will trigger cluster formation phase again and then dynamically re-select CHs according to the change of node residual energy, otherwise, it will only perform the cluster selection phase.

Additionally, in order to verify the effectiveness of our cluster formation algorithm, we also combined the FCM-based cluster formation algorithm which was used in [20] with our proposed CH selection algorithm to form the clustering routing protocol based FCM (CRPFCM).

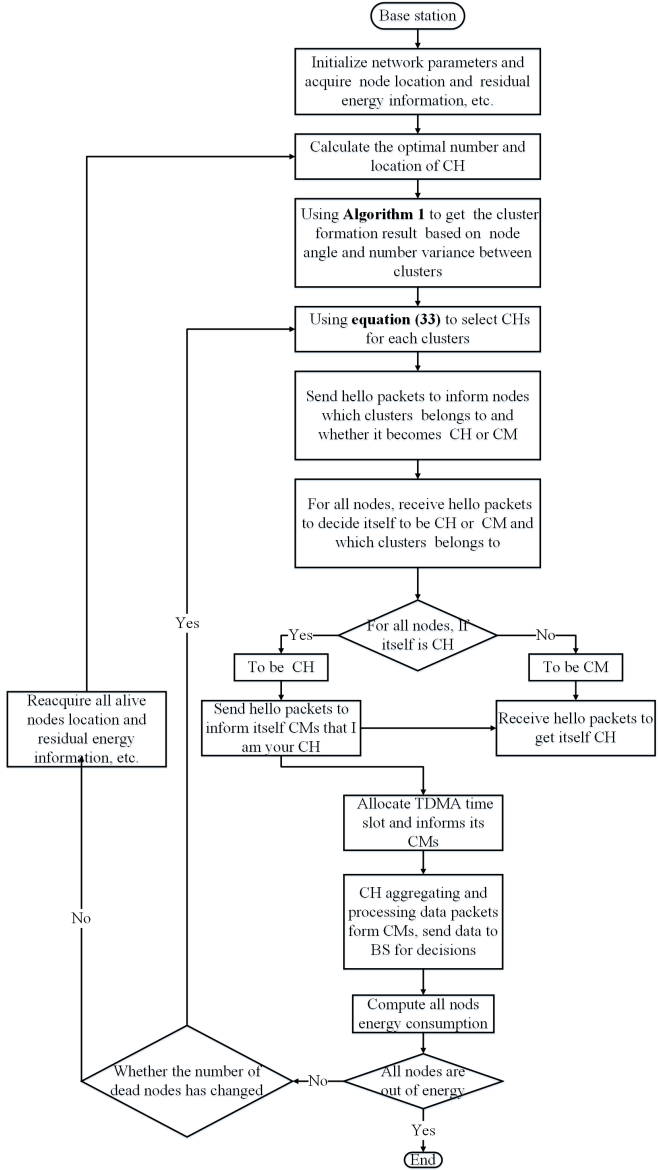

Fig. 5. Workflow of EERPMS.

Next, we will introduce the three phases of the EERPMS protocol respectively. Before introducing the phase of cluster formation, we first derive the closed-form expressions of the optimal number and location of CHs which will be used in the phases of CH selection and cluster formation.

### A. Theory of Optimal CH Location and Number

The location and number of CHs have a great influence on data transmission efficiency and network energy consumption. With the goal of minimizing the network energy consumption, we put forward *Theorem 2* to determine the optimal number of CHs ($K^*$) and the optimal distance between CH and BS ($d^*_{CHtoBS}$).

*Theorem 2* When $K$ and $d_{CHtoBS}$ meet the conditions shown in (15) and (16), respectively, the total network energy consumption can be minimized.

$$d^*_{CHtoBS} = \frac{2NR}{3(N+K^*)} \quad (15)$$

$$K^* = (\frac{3}{4}\pi^2 N)^{\frac{1}{3}} \quad (16)$$

Based on the optimal distance between CH and BS ($d^*_{CHtoBS}$), we define the minimum energy consumption circle $O_1$ with BS as the center and radius $R_{O_1} = d^*_{CHtoBS}$.

*Proof:* Assume that there are $N$ nodes distributed uniformly in the circular area with radius $R$. If the area is divided into $K$ clusters and each cluster includes one CH and $N/K-1$ CMs, thus the energy consumption of each clusters is composed of one CH energy consumption and $N/K-1$ CMs energy consumption. The CH energy consumption mainly comes from the following three parts:

1) to receive the data packets from $N/K-1$ CMs.
2) to fuse the data packets from $N/K-1$ CMs and itself.
3) to send the data packets to BS.

Thanks to the condition of *Theorem 1*, the energy consumption can be optimized and follows the free-space model ($d^2$ power loss). So the CH energy consumption can be described in (17).

$$E_{CHtoBS} = lE_{elec}(\frac{N}{K}-1) + lE_{DA}\frac{N}{K} + (lE_{elec} + l\varepsilon_{fs}d^2_{CHtoBS}) \quad (17)$$

where, $E_{DA}$ is the energy consumption of fusing 1-bit data packet.

CM's energy is consumed by sending their data to the CH. Generally, CMs are close to the CH, presumably the distance to CH is small, so the energy consumption follows the free-space model ($d^2$ power loss). Thus, the energy used in CMs is calculated by (18).

$$E_{CMtoCH} = lE_{elec} + l\varepsilon_{fs}d^2_{CMtoCH} \quad (18)$$

where $d_{CMtoCH}$ is the distance between CM and CH. The area occupied by each cluster is approximately $(\pi R^2)/K$. Generally, nodes are uniformly distributed in an arbitrarily shaped region with a density function $\rho(x, y)$. We assume that the region is a sector (apex angle for $\gamma = 2\pi/K$), and CH is located on the angular bisector of the sector, as shown in Fig. 2. Therefore, density function $\rho(x, y) = K/(\pi R^2)$ and the expected quadratic distance between CMs and CH is shown in (19).

$$E[d^2_{CMtoCH}] = \int_0^R \int_{-xtan(\pi/K)}^{xtan(\pi/K)} [(x - d_{CHtoBS})^2 + y^2]\rho \, dy dx$$

$$= d^2_{CHtoBS} - \frac{4R}{3}d_{CHtoBS} + \frac{R^2}{2} + \frac{\pi^2 R^2}{6K^2} \quad (19)$$

*Remark 1:* Note that although the shape of monitoring area and cluster are assumed to be circular and fan-shaped, our theoretical derivation process is suitable for monitoring areas and clusters with any shape.

Based on (17) and (18), the energy consumption of each cluster and the total network energy consumption can be calculated as in (20) and (21), respectively.

$$E_{Cluster} = E_{CHtoBS} + (\frac{N}{K} - 1) \times E_{CMtoCH}$$
$$\approx E_{CHtoBS} + \frac{N}{K} \times E_{CMtoCH}$$
$$= \frac{l}{K}[2E_{elce}N + E_{DA}N + K\varepsilon_{fs}d_{CHtoBS}^2 + \ldots$$
$$+ N\varepsilon_{fs}d_{CMtoCH}^2] \quad (20)$$
$$E_{Total} = KE_{Cluster}$$
$$= l[\varepsilon_{fs}((K+N)d_{CHtoBS}^2 - \frac{4NR}{3}d_{CHtoBS}) + \ldots$$
$$+ N(2E_{elec} + E_{DA} + \frac{\varepsilon_{fs}(3K^2 + \pi^2)R^2}{6K^2})] \quad (21)$$

Obviously, the total network energy consumption $E_{Total}$ can be minimized by method for finding extremum of multivariate function (setting the derivative of $E_{Total}$ with respect to $K$ and $d_{CHtoBS}$ to zero), so $K^*$ and $d^*_{CHtoBS}$ is obtained. The proof is completed.

### B. Cluster Formation

In this part, the mapping relationship between multi-threshold image segmentation and cluster formation is constructed at first, and then node clustering is completed by multi-threshold Otsu algorithm.

*1) The Mapping Between Multi-Threshold Image Segmentation and Cluster Formation:* The principle of multi-threshold image segmentation [27] is to segment the gray level of an image by selecting the appropriate segmentation threshold according to the difference of gray value, so that the gray value shows consistency or similarity in the same area, while obvious differences across different areas. The purpose of node clustering is to ensure that some attributes of nodes, such as coordinates and angles of nodes, are very similar in the same cluster, but there are great differences in different clusters, which is consistent with multi-threshold image segmentation. Therefore, we try to apply the multi-threshold image segmentation algorithm to clustering nodes.

When we use the angle attribute of nodes to cluster nodes, the mapping with multi-threshold image segmentation is shown in Fig. 6. An image to be segmented is analogous to a monitored area to be clustered, and reading the gray value of the image is similar to calculating the angle of nodes (the angle calculation of nodes takes BS as a reference). Therefore, the image segmentation using the grayscale threshold is equivalent to the node clustering using the angular threshold.

The motivation for selecting the angle attribute of nodes to clustering nodes is as follows. By grouping the nodes with similar angles into the same cluster, the nodes in the same cluster will be limited to a certain angle range, thus avoiding the situation that the distance between the nodes in the cluster varies greatly. As illustrated in Fig. 7, the angle values of $P_1$ and $P_2$ are quite different, while the angle values difference of $P_2$ and $P_3$ are small. So we can conclude that the distance between $P_2$ and $P_3$ is smaller, which is more suitable for being divided into the same cluster. Therefore, the benefit of selecting the angle attribute of nodes to clustering nodes is to shorten the communication distance between nodes in the same cluster, namely, to reduce the energy consumption between CH and CMs.

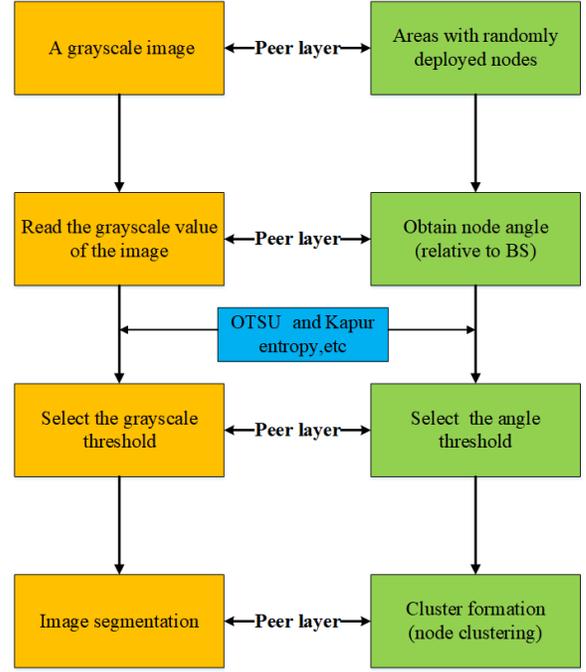

Fig. 6. Mapping of multi-threshold image segmentation and node clustering.

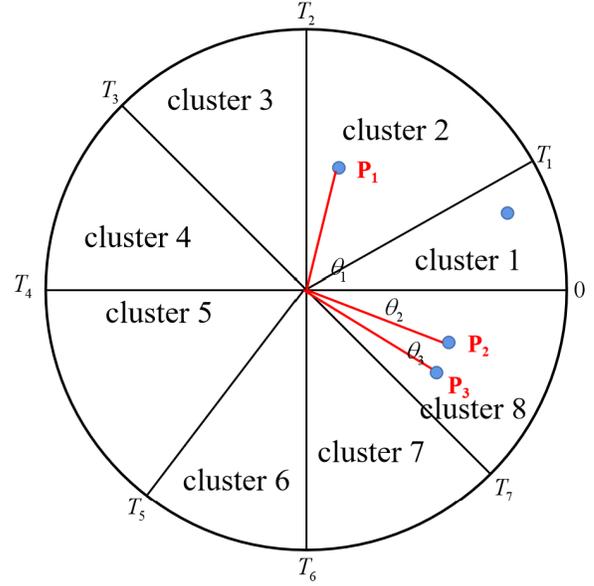

Fig. 7. Schematic diagram of node clustering.

*2) Node Clustering Based on Variance of Node Angles and Number of Nodes Between Clusters:* Based on the above analysis, the problem of node clustering is transformed into the problem of how to find the best angle segmentation threshold, so as to realize the optimal clustering of nodes. Among many multi-threshold segmentation algorithms, the Otsu algorithm has been widely used because of its simplicity, high efficiency, and parameters free [28]. Therefore, this paper divides nodes into optimal clusters by integrating with Otsu, angle and number of nodes.

The principle of node clustering is as follows. There are $N$ sensor nodes deployed in the monitoring area, which needs to be divided into $K^*$ clusters. $C$ represents the cluster set, and $C = \{C_1, C_2, \ldots, C_{K^*}\}$. The angular range of nodes contained in each cluster is denoted by $M = \{M_1, M_2, \ldots, M_{K^*}\}$. Where $M_1 \in (0, T_1 - 1]$, $M_2 \in [T_1, T_2 - 1], \ldots, M_{K^*} \in [T^*_{K-1}, 2\pi]$. $T$ indicate the angle segmentation threshold set and $T = \{T_1, T_2, \ldots, T_{K^*-1}\}$.

To obtain optimal threshold $T^*$, the probability of occurrence of each angle value is firstly calculated by $p_i = \frac{n_{\theta_i}}{N}$ ($n_{\theta_i}$ represents the occurrence frequency of $\theta_i$), and then the node angle mean of each cluster $u_k$ can be calculated by (22). $w_1, w_2, \ldots, w_{K^*}$ represent the cumulative probability of angle values of each cluster respectively, as shown in (23). Finally, the expression of the angle variance of inter-clusters nodes $f_1(T)$ is shown in (24).

$$u_k = \sum_{\theta_i \in M_k} \theta_i p_i / w_k, k = 1, 2, \ldots, K^* \quad (22)$$

$$w_k = \sum_{\theta_i \in M_k} p_i, k = 1, 2, \ldots, K^* \quad (23)$$

$$f_1(T) = \sum_{k=1}^{K^*} w_k (u_k - u_T)^2 \quad (24)$$

where, $u_T$ represents the mean value of angle values of all nodes in the whole region. The calculation is shown in (25):

$$u_T = \sum_{k=1}^{K^*} w_k u_k \quad (25)$$

After the node clustering is obtained based on the angle segmentation threshold $(T_1, T_2, T_3, \ldots T^*_{K-1})$, the number of nodes in each cluster $(NC_k, k \in [1, K^*])$ can also be obtained, and then the number variance of inter-cluster nodes can be calculated by (26). which is beneficial to CH load balance. Note that the smaller the value of $f_2$ is, the better the cluster head load balance will be.

$$f_2(T) = \frac{1}{N} \sum_{k=1}^{K^*} (NC_k - NC_T)^2 \quad (26)$$

where $NC_T$ represents the average number of nodes and $NC_T = N/K^*$.

Based on (24) and (26), $f_1$ and $f_2$ are normalized and linearly weighted to obtain the objective function $F_1$. $\overline{f_1}$ and $\overline{f_2}$ respectively represent normalized functions. Note that if the objective function $F_1$ is maximized, nodes with similar angles will be divided into the same cluster, and the number of nodes in each cluster is approximately equal. Namely, the number of CMs for each cluster head service (CH load balance) is approximately equal and the uniform distribution of CH can be guaranteed because CHs are distributed in every angular direction of the monitoring area (see Fig. 7).

$$F_1 = \alpha_1 \overline{f_1} + \alpha_2 / \overline{f_2} \quad (27)$$

Here, $\alpha_1$ and $\alpha_2$ are the control parameteres in the range [0,1], with $\alpha_1 + \alpha_2 = 1$.

*3) Acquire Optimal Clusters:* Since finding optimal clusters is a NP-hard problem, an optimization algorithm needs to be exploited to optimize the optimal clusters. SI algorithms such as particle swarm algorithm, artificial bee colony and Bat Algorithm (BA) have been applied to find the optimal segmentation threshold and have achieved excellent performance [29]. In view of the fact that these SI algorithms have almost no difference in the accuracy of solving the optimal solution of $F_1$, we adopt BA [30] to acquire $T^*$ of objective function.

In the process of solving the optimal segmentation threshold $T^*$ using BA, the bat population is expressed as $x = (T_1, T_2, \ldots, T_{K^*-1})$. (28) and (29) are used to update the velocity $v_i^t$ and position $x_i^t$ of the $i_{th}$ bat at time $t$.

$$v_i^{t+1} = v_i^t + (x_i^t - x^*) \times S_i \quad (28)$$
$$x_i^{t+1} = \lceil x_i^t + v_i^{t+1} \rceil \quad (29)$$

The frequency of bats is updated by (30).

$$S_i = S_{min} + (S_{max} - S_{min}) \times \beta \quad (30)$$

where, $S_{min}$ and $S_{max}$ represent the minimum and maximum frequency of bats, respectively. $\beta$ is a uniformly distributed random vector with $\beta \in [0, 1]$.

When one solution is accepted as the current optimal solution, the new solution is locally updated by (31).

$$x_i^{t+1} = \lceil x_i^t + \varepsilon \overline{A}^t \rceil \quad (31)$$

where, $\varepsilon \in [0, 1]$ is a uniformly distributed random vector. $x^*$ is the global optimal solution. $\overline{A}^t$ denotes average loudness of bats. $\lceil \rceil$ stands for rounding up. The loudness and frequency updating methods of bats follow (26) and (27) respectively. $\epsilon$ and $\gamma$ are the constant.

$$A_i^{t+1} = \epsilon A_i^t \quad (32)$$
$$r_i^{t+1} = r_i^0 [1 - \exp(-\gamma t)] \quad (33)$$

The pseudo-code of the cluster formation is shown in **Algorithm 1**.

### C. CH Selection

After the cluster formation phase is over, the appropriate CH should be selected in each cluster. According to *Theorem 2*, when CHs are distributed on $O_1$, the total network energy consumption can be minimized. Therefore, the selected CHs should near $O_1$ as much as possible. At the same time, nodes with more residual energy should act as CHs so as to prevent the nodes near $O_1$ acting as CHs frequently and dying soon. Therefore, an attribute function [19] related to optimal CH location and node residual energy is designed as shown in (34). Then, node $s_i$ with the maximum attribute value is selected as the CH in each cluster.

$$F_2(s_i) = \omega_1 \frac{E_{re}(s_i)}{E_o(s_i)} + \omega_2 \frac{d_{maxtoO_1} - d_{s_i toO_1}}{d_{maxtoO_1} - d_{mintoO_1}} \quad (34)$$

where, $\omega_1$ and $\omega_2$ are the control parameteres in the range [0,1], with $\omega_1 + \omega_2 = 1$. $d_{maxtoO_1}$ and $d_{mintoO_1}$ represent

**Algorithm 1** Cluster Formation Algorithm

**Require:** $N$ sensor nodes; Clusters number $K^*$;
**Ensure:** Angle segmentation threshold set $T_{opt}$; Optimal clusters set $C_{opt}$
1: Initialize bat population $x_i$, and velocity $v_i$;
2: Define frequency $f_i$;
3: Initial pulse emission rate r and loudness A;
4: **repeat**
5:    Generate new solutions by adjusting frequency and updating velocity and location by (22) to (23);
6:    **if** $rand > r_i$ **then**
7:       Select a solution among best solutions;
8:       Generate new local solution around selected best solution using (21);
9:    **end if**
10:   **for** sensor nodes $s_i, i = 1 \to N$ **do**
11:       Calculate the angle $\theta_i$ of node $s_i$;
12:   **end for**
13:   Count the frequency $n_{\theta_i}$ of each angle $\theta_i$;
14:   Calculate the probability $p_i$ of $\theta_i$;
15:   acquire clusters set $C_k, k = 1 \to K^*$
16:   **for** clusters set $C_k, k = 1 \to K^*$ **do**
17:       Calculate the $u_k$ in $C_k$ using (16);
18:       Calculate the $w_k$ in $C_k$ using (17);
19:       Calculate the $NC_k$ in $C_k$;
20:   **end for**
21:   Compute $u_T$ using (19);
22:   Compute $f_1$ using (16) and normalize $\overline{f_1}$;
23:   Compute $f_2$ using (20) and normalize $\overline{f_2}$;
24:   Calculate the objective function value using (21);
25:   **if** $rand < A_i$ **then** and $F_1(x_i) < F_1(x^*)$
26:       Accept the new solution;
27:       Decrease $A_i$, increase $r_i$, by (26) and (27);
28:   **end if**
29:   Rank the bats and find the current best $x^*$;
30: **until** $t < t_{max}$
31: Let $T^* = x^*$ and acquire $C_{opt}$;

the maximum and minimum distance intra-cluster all nodes to $O_1$, respectively. $d_{s_i to O_1}$ represents the distance from $s_i$ to $O_1$. $E_{re}(s_i)$ and $E_o(s_i)$ represent the residual and initial energy of $s_i$, respectively.

### D. Data Transmission

Once CHs are selected and the transmission scheduled is made, CMs will periodically collect soil moisture data and transmit the data to CH in their TDMA timeslot. Then, CHs receive and aggregate this data and forwards it to BS.

## V. SIMULATION AND ANALYSIS

This section provides the experimental results to evaluate the proposed EERPMS protocol's efficiency for reducing network energy consumption and extending network lifetime. Firstly, the parameter setting of EERPMS is discussed. Next, some performance indicators including the CH distribution, CH energy consumption, load balance, network lifetime and energy consumption are compared with RLEACH [17], FIGWO [19] and CRPFCM (see section IV. D). All the routing protocols were coded and executed on MATLAB R2016b on Windows 10-64 bit on a PC with Intel CoreI7.

TABLE II
COMPARISONS OF FIVE PROTOCOLS

| Protocols | Cluster formation algorithm | CH selection algorithm (Considerations) |
|---|---|---|
| RLEACH (LEACH-based) | Join the nearest CH | Probability Election (Node residual energy) |
| FIGWO (SI-based) | Join the nearest CH | Objective Function (Node residual energy; Distance form CH to BS) |
| CRPFCM (AI-based) | Fuzzy C-means | Attribute Function (node residual energy; distance form CH to $O_1$) |
| EERPMS (Multi-threshold segmentation-based) | Inter-clusters variance of node angle and number | Attribute Function (Node residual energy; Distance form CH to $O_1$) |
| TS-RBHR (Region-based) | Static region division | Fuzzy Logic (Node residual energy; Distance form CH to BS) |

LEACH-based, swarm intelligence-based and AI-based clustering methods are often used by researchers to form clustering routing protocols. We select three representative algorithms from each of the above methods where RLEACH is LEACH-based protocol, FIGWO is swarm intelligence-based and CRPFCM is AI-based. Of course, another reason for choosing these comparison protocols is that the three comparison protocols are similar to our proposed protocol in some dimensions (refer Table II). By comparing with them, the effectiveness of our proposed protocol for cluster formation and cluster head selection could be verified. Lastly, we compare EERPMS with TS-RBHR (Hybrid Routing Protocol Based on Threshold Sensitive Region) [14], which is a region-based static routing protocol for precision agriculture.

### A. Theoretical Verification of Optimal CH Location and Number

Based on *Theorem 2*, the theoretical values of $d^*_{CHtoBS}$ and $K^*$ are 91.74 m and 9, respectively (when $R = 150m$ and $N = 100$). In order to evaluate the correctness of *Theorem 2* and provide optimal parameters for subsequent simulations, $\omega_1$ and $\omega_2$ in (34) are taken as 0 and 1, respectively (maximizing the effect of the optimal location and number of CHs). In Fig. 8, we can clearly conclude that when $K = 1$, the number of segmentation threshold is 0, which is equivalent to BS acting a CH. It's the same thing as having no CH ($K = 0$). In this case, the energy consumption of both networks is the same. When $K$ remains unchanged and $K > 1$, with the increase in $d_{CHtoBS}$, the network energy consumption shows a trend of first declining and then rising. When $d_{CHtoBS}$ remains unchanged, and as $K$ increases, the network energy consumption also shows a trend of first declining and then rising. When $K \approx 10$ and $d_{CHtoBS} \approx 90m$, the network energy consumption is the lowest. The theoretical analysis limits the uniformity of node distribution and the shape of the cluster area, which are difficult to meet at any time in the experiment. Therefore, there is a slight difference between the experimental value and the theoretical value within the acceptable range. This experimental result verifies the correctness of the proposed *Theorem 2*.

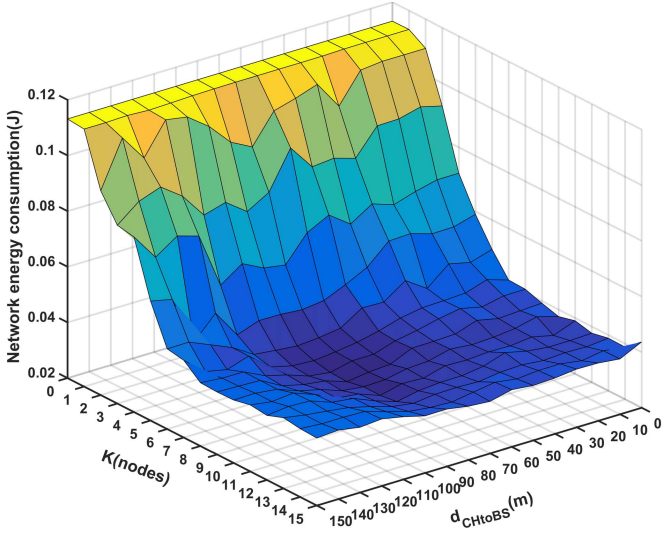

Fig. 8. Influence of the location and number of CHs on network energy consumption, when $R = 150m$ and $N = 100$.

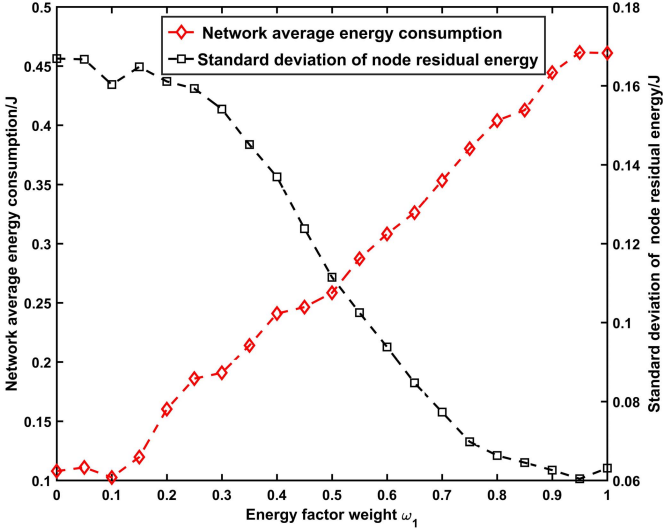

Fig. 9. Influence of $\omega_1$ on network consumption and standard deviation of node residual energy, when $R = 150m$ and $N = 100$.

### B. Influence of Distance and Energy Weight on Network

As discussed in Section IV, the CH's location and residual energy have a great influence on network lifetime and energy consumption. Therefore, it is necessary to analyze the influence of the proportion of $\omega_1$ and $\omega_2$ on the network. Based on the different $\omega_1$, this section makes statistics on the average energy consumption, standard deviation of node residual energy, and the rounds where the first and last node die (see Figs. 9 and 10).

In Fig. 9, when $\omega_1$ continues to increase, the network average energy consumption presents an upward trend, while the standard deviation of node residual energy presents a downward trend. This is because when $\omega_1$ is larger, most CHs are not near $O_1$, which leads to the increase of network energy consumption. Meanwhile, due to the increase of $\omega_1$, nodes with more residual energy act as CHs, avoiding the frequent use of nodes near $O_1$ as CHs; this results in energy consumption to be averaged by more nodes. Therefore, the standard deviation of residual energy is reduced, that is, the energy consumption between nodes is more balanced.

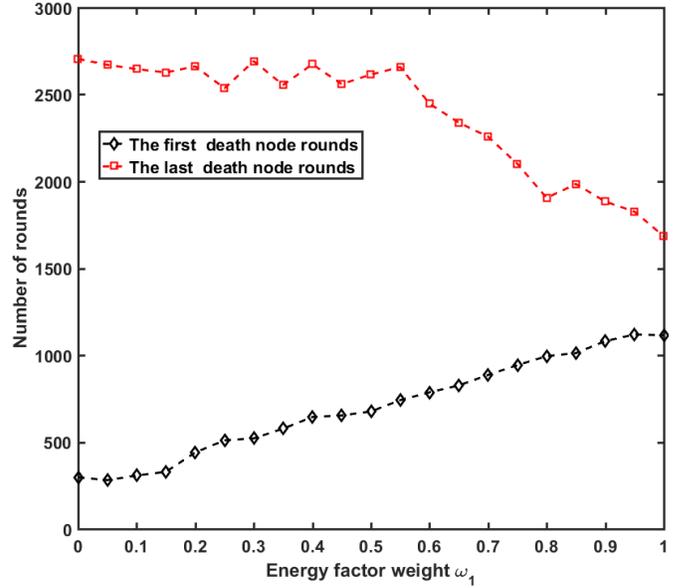

Fig. 10. Influence of $\omega_1$ on death node round.

In Fig. 10, the first node death round increases, and the number of rounds of death of all nodes decreases with the increase in $\omega_1$. This is because CH undertake a large number of relay and forwarding tasks, which leads to a high energy consumption, and thus CHs is most likely to die. Therefore, the first node death round is mainly affected by CHs. When $\omega_1$ is increased, frequent acting of CHs near $O_1$ is avoided, and the number of CHs available is increased, thus extending each CH's survival time; that is, the death time of the first node is prolonged. Furthermore, the death round of all nodes is affected by the overall energy consumption of the network. As the energy factor $\omega_1$ increase, CH is far away from $O_1$, and the whole network consumes a lot of energy, leading to the premature death of all nodes in the network. However, different values of $\omega_1$ has different functions. We should determine the value of $\omega_1$ according to scenario requirements. Specifically, in pursuit of longer last death node round (LDN), we can set $\omega_1 = 1$. while $\omega_1$ can be set to 0 for the long first death node round (FDN). Therefore, motivating by the tradeoff between FDN and LDN and giving consideration to network energy consumption balance, the following simulation will take $\omega_1 = 0.7$ as an example for further discussion. The parameters involved in the following comparative experiments are shown in Table III.

### C. CH Distribution and Number

CH distribution has great impact on network performance. The more uniform CHs are distributed, the smaller and more balanced the energy consumption between nodes is. We simulated an area of $R = 150\ m$ with randomly deploying 100 nodes to compare CH distribution among RLEACH, FIGWO, CRPFCM and EERPMS. Fig. 11 clearly shows that the distribution of CHs in RLEACH is very nonuniform and random, and there is a great difference in the number of intra-cluster CMs. This is caused by the probability-based CH selection method and the mode that nodes join CH nearby to form clusters. In addition, due to the nonuniform distribution of CHs, the distance between CMs and CH and the distance between CHs and BS are very large.

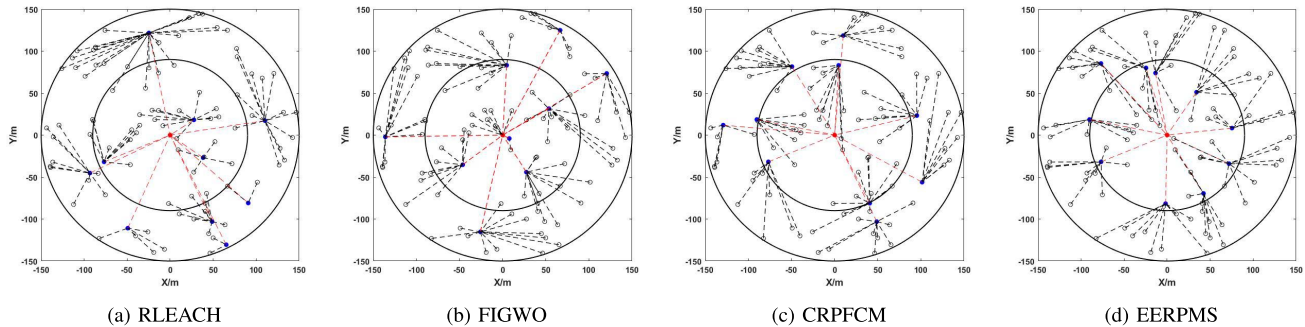

(a) RLEACH  (b) FIGWO  (c) CRPFCM  (d) EERPMS

Fig. 11. Cluster formation and CH distribution at arbitrary round, when $R = 150m$ and $N = 100$. The blue and red solid dots represent the CHs and BS, respectively, and black dotted lines represent the virtual transmission path of CMs. The inner circle represents $O_1$.

TABLE III
EXPERIMENTAL PARAMETERS

| Parameters | Values |
|---|---|
| $R$ | $150\ m$ |
| $N$ | 50-250 nodes |
| $E_o$ | $0.5\ J$ |
| $l$ | $4000\ bits$ |
| $E_{mp}$ | $0.0013\ pJ/bit/m^4$ |
| $E_{fs}$ | $10\ pJ/bit/m^2$ |
| $E_{elec}$ | $50\ nJ/bit$ |
| $E_{DA}$ | $5\ nJ/bit$ |
| $\omega_1, \omega_2$ | 0.7, 0.3 |
| $\alpha_1, \alpha_2$ | 0.5, 0.5 |
| $K^*$ | 10 nodes |
| $R_{O_1}$ | $90\ m$ |

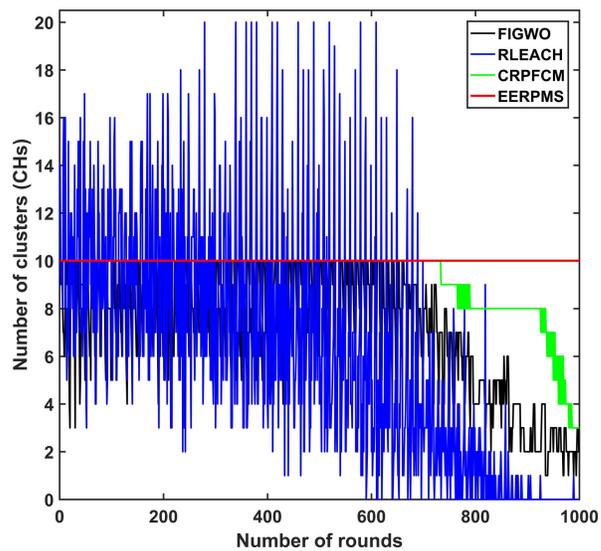

Fig. 12. The number of CHs in the first 1000 rounds.

For FIGWO, CH distribution is also nonuniform. Worse, there is no CH at all in some areas, and many CHs are very close to or far away from BS, resulting in a very long communication distance between CH and CMs. The reason is that the prey position determined by the first three wolves is often an area where there is no node. Finding the node closest to the prey is selected as CH, which easily leads to the node with lower fitness value being selected as CH, thus the uniformity of CH distribution cannot be guaranteed.

For CRPFCM, the distance between CMs and CH is small because the principle of cluster formation based on FCM can assist to minimize the total spatial distance between CH and CMs. However, many clusters are far away from $O_1$, so there is no guarantee that more CHs will be distributed near $O_1$.

For EERPMS, CHs are uniformly distributed in the monitored area. In the same cluster, the distance between nodes is very small in the vertical direction, which is the advantage of maximizing the angular variance of nodes between clusters. Although the distance between nodes in the horizontal direction may be large, in our analysis scenario (see *Theorem 2*), the existence of CHs can ensure that the communication distance between nodes is limited within $d_{th}$, thus reducing network energy consumption. Secondly, the clustering method based on minimizing the variance number of nodes between clusters makes the size of each cluster uniform, which is in favor of CHs load balance. Lastly, all CHs are distributed near in $O_1$, which can greatly contribute to reducing network energy consumption.

According to *Theorem 2*, CH number is closely related to the network energy consumption. As is demonstrated in Fig. 12, for RLEACH, the method of CH selection based on probability and node rotation will result in the CH number deviating from the optimal CH number and sorely fluctuation. For FIGWO, if the number of CMs in a cluster is less than three, this kind of cluster will not select CH. Thus the CH number will be less than the optimal value in most cases. For EERPMS and CRPFCM, as long as the number of surviving nodes does not change, the number of clusters will always remain at the optimal number. Hence, before the death node appears, the number of clusters can be maintained at 10.

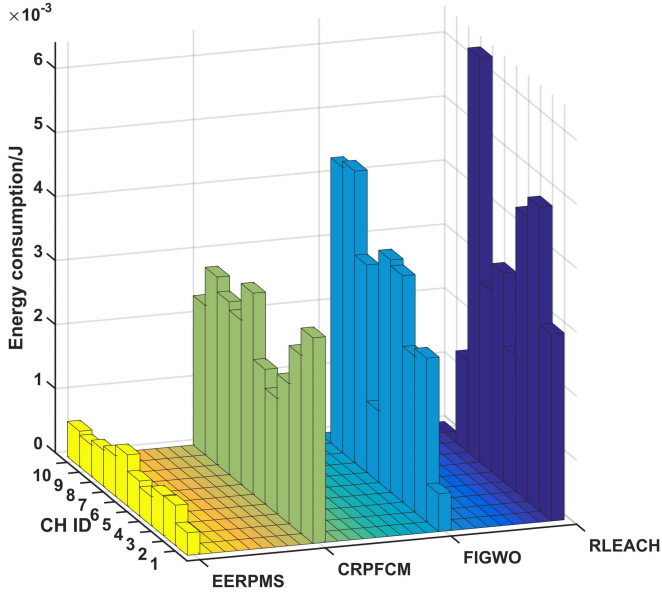

Fig. 13. The energy consumption of each CH is compared in the 100th round, when $R = 150m$ and $N = 100$.

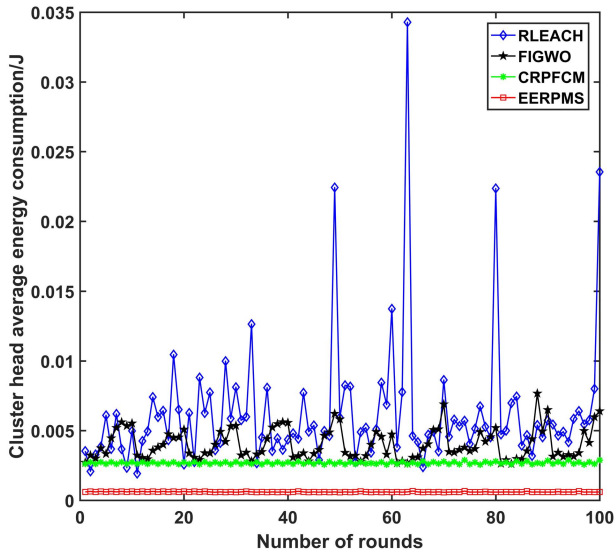

Fig. 14. Comparison of CH energy consumption at 100th round, when $R = 150m$ and $N = 100$.

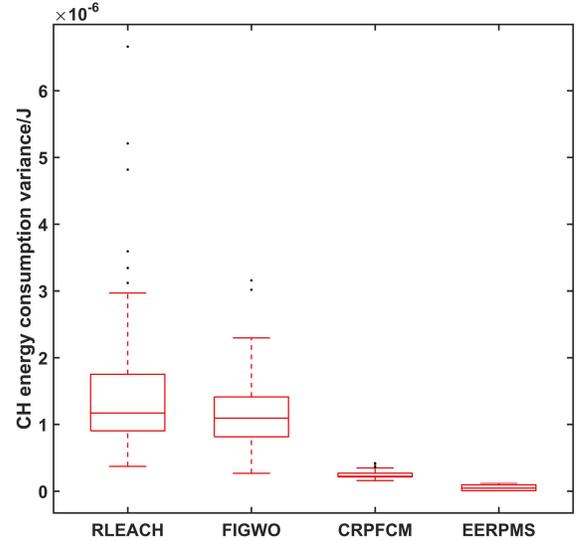

Fig. 15. Comparison of CH load balance, when $R = 150m$ and $N = 100$.

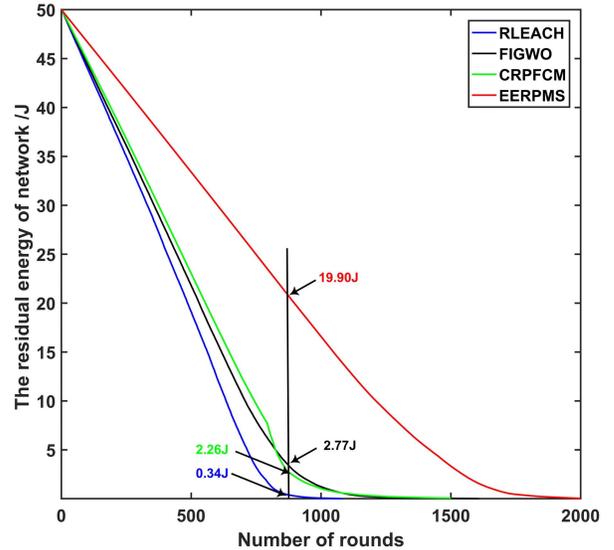

Fig. 16. Residual energy of network, when $R = 150m$ and $N = 100$.

### D. CH Energy Consumption and Load Balance

CHs usually consume much more energy than CMs and thus are more likely to run out of energy early. Balanced load and low CH energy consumption can prolong the CH working time and network lifetime.

Fig. 14 shows that the energy consumption of each CH is compared in the 100th round. It can be concluded that CHs energy consumption of EERPMS protocol is much lower than that of the other three algorithms, and the energy consumption difference between CHs is smaller (load balance is better). In order to further evaluate CH energy consumption and load balance, we compared the performance of four algorithms in the first 100 rounds. In Fig. 14, RLEACH's CH consumes the most energy and fluctuates greatly, and FIGWO's CH also consumes a lot of energy and fluctuates. The energy consumption of our proposed EERPMS has always kept at the lowest level. We use the variance of CH energy consumption in a round to measure the load balance of CHs. Because the load balance factor is considered in clustering (minimize the variance of node number in clustering nodes), the number of loads carried by CHs in each round is basically the same. Therefore, the variance of CH energy consumption is the smallest in each round (the CH load balance is better), as shown in Fig. 15.

### E. Network Energy Consumption and Network Lifetime

The network residual energy refers to the sum of all nodes' residual energy, which is a key factor to measure the network energy utilization efficiency. Fig. 16 gives the comparison of network residual energy among RLEACH, FIGWO, CRPFCM and EERPMS. The network residual energy of EERPMS is the most, that is, the network energy consumption of our proposed EERPMS is the lowest. Especially, EERPMS can save up to 64.50%, 58.60% and 56.15% network energy at 800th round as compared to RLEACH, CRPFCM and FIGWO protocols respectively. This is mainly due to the optimal CH location theorem aiming at network energy consumption. Although CRPFCM is based on the above theorem when selecting CHs, clustering results based on FCM will cause some clusters to be far away from $O_1$, so it cannot effectively reduce network energy consumption. For RLEACH, the location and number of CHs are very random, which makes it more difficult to meet the conditions in *Theorem 2*. For FIGWO, it stipulates that if there are fewer than 3 CMs, CH will be regarded as ordinary nodes, so in most cases, the number of cluster heads is less than the optimal value, and the nodes closer to BS are given priority when CH is selected, so it is also difficult to satisfy the conditions in *Theorem 2*. In Fig. 17, the node energy consumption in the 100th round is shown. We can see that the node energy consumption of the proposed EERPMS is much lower than other algorithms. Based on the simulation results above, the proposed EERPMS can effectively reduce the energy consumption of nodes.

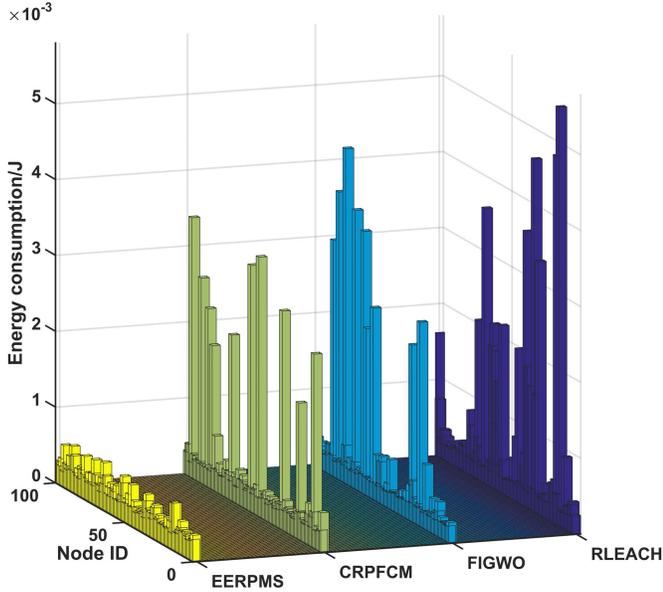

Fig. 17. Comparison of node energy consumption at 100th round, when $R = 150m$ and $N = 100$.

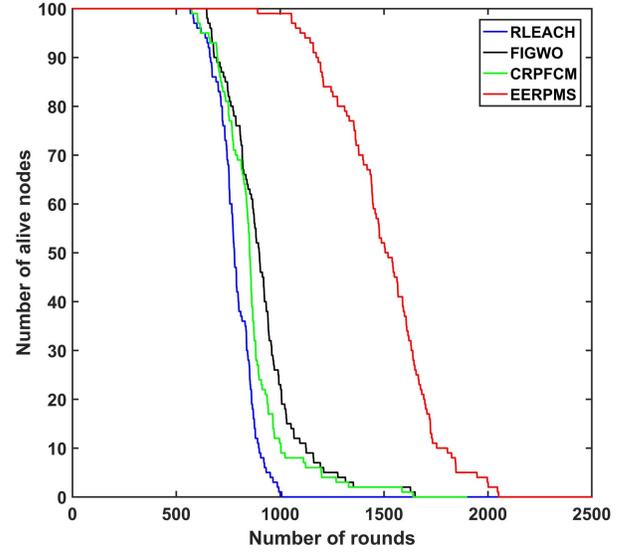

Fig. 18. Number of nodes alive versus the number of rounds, when $R = 150m$ and $N = 100$.

Network lifetime can be reflected by the round in which the dead node is located. With regard to network lifetime, the most popular definitions are the first dead node (FDN), half dead nodes (HDN) and last dead node (LDN). Network energy consumption is the most important factor that affects HDN and LDN. In the case of the same total network energy, if the network's energy consumption in each round is lower, HDN and LDN will be longer. CHs take on more forwarding tasks and consume more energy than CMs, so FDN is mainly affected by the death round of CHs.

Fig. 18 reflects the changes of surviving nodes with the network operation. It can be clearly seen that the number of surviving nodes of EERPMS is always the largest, so compared with RLAECH, FIGWO and CRPFCM, the network lifetime of EERPMS is always the largest due to the lowest network energy consumption. In order to eliminate the influence of experimental contingency, we conducted 100 random and independent experiments, and counted the network lifetime of four algorithms in Table IV. It can be concluded that EERPMS has increased by 57.04%, 55.67% and 51.70% in FDN compared with RLEACH, FIGWO and CRPFCM respectively. EERPMS increased by 89.35%, 71.33% and 83.80% in HDN and 103.83%, 23.04% and 28.69% in LDN, respectively.

TABLE IV
COMPARISON OF NETWORK LIFETIME

| Network Lifetime | RLEACH | FIGWO | CRPFCM | EERPMS |
|---|---|---|---|---|
| FDN | 568 | 573 | 588 | 892 |
| Improvements | 57.04% | 55.67% | 51.70% | – |
| HDN | 827 | 914 | 852 | 1566 |
| Improvements | 89.35% | 71.33% | 80.30% | – |
| LDN | 1019 | 1688 | 1614 | 2077 |
| Improvements | 103.83% | 23.04% | 28.69% | – |

In view of the monitoring area conforming to *Theorem 1*, the varies of monitoring area size has little effect on the network lifetime, so we only discuss the effect of node density on network lifetime and conduct different experiments under the case that $R = 150m$ and the number of nodes varies from 50 to 250. Figs. 19-21 clearly shows that the proposed EERPMS protocol outperforms other three protocols for all case of the node density, FDN, HDN and LDN. All four protocols display their shortest network lifetimes (FDN, HDN and LDN) simultaneously when the number of nodes is the minimum value 50. This is because sensor nodes become CHs more frequently for a network with lower node density. However, for EERPMS, with the increase of node density, the number of nodes near $O_1$ increases, and the number of available CHs increases, which more remarkably reduces the network energy consumption and node energy consumption rate. Hence, EERPMS has a longer network lifetime.

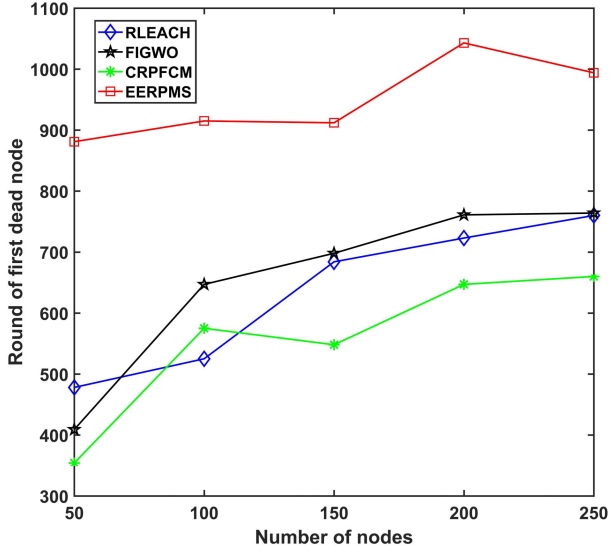

Fig. 19. Round of first dead nodes versus the different number of nodes, when $R = 150m$ and the number of nodes varies from 50 to 250.

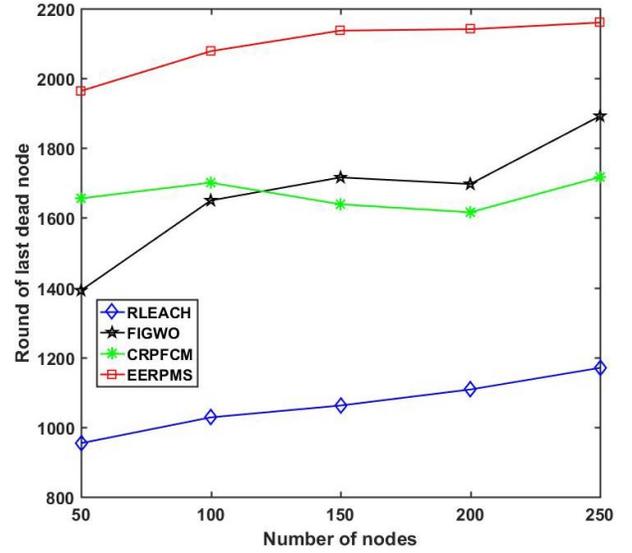

Fig. 21. Round of last dead nodes versus the different number of nodes, when $R = 150m$ and the number of nodes varies from 50 to 250.

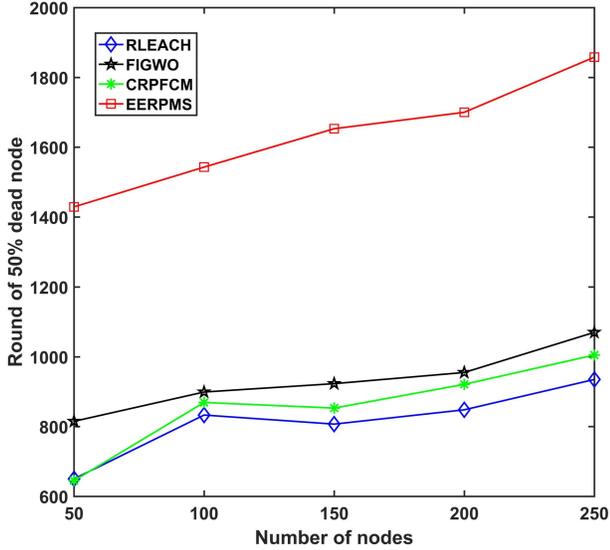

Fig. 20. Round of 50% dead nodes versus the different number of nodes, when $R = 150m$ and the number of nodes varies from 50 to 250.

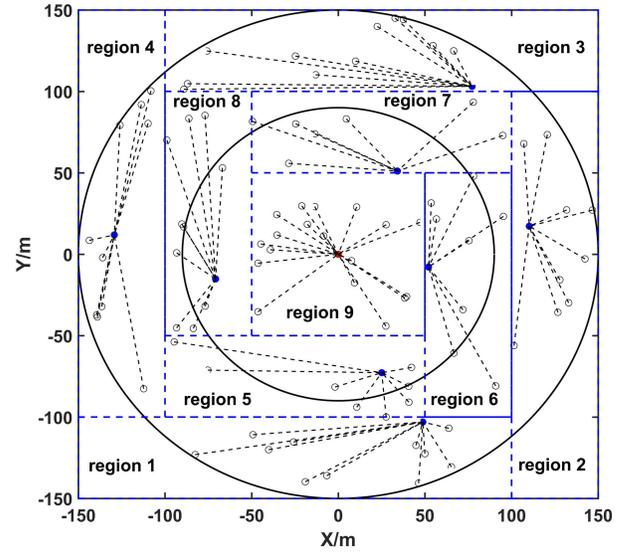

Fig. 22. The area division and data transmission path of TS-RBHR protocol.

### F. EERPMS Versus TS-RBHR

First, we simulate a square area of $150m \times 150m$ for TS-RBHR protocol, which has 100 nodes (the number of the Type-1 and Type-2 node are 40 and 60, respectively) and is divided into 9 sub-regions (see Fig. 22). All Type-1 nodes are deployed in region 9 and Type-2 nodes are in regions 1-8. The initial energy of the Type-1 node and Type-2 node is $0.2J$ and $0.566J$, respectively. $TH_1$ (minimum threshold for "node's distance From BS") and $TH_2$ (maximum threshold for "node's distance From BS") is set as $80m$ and $120m$. $TH_3$ (minimum threshold for "residual energy") and $TH_4$ (maximum threshold for "residual energy") is set as $0.2J$ and $0.4J$. Note that for the fairness of comparison, we default that TS-RBHR will meet the threshold condition in each round.

In Fig. 23, the number of alive nodes varying with the network energy consumption is reflected. EERPMS always has more alive nodes than TS-RBHR. Note that when the network consumes about $30J$ energy, the number of alive nodes of EERPMS and TS-RBHR is 100 and 67 respectively. Hence, we can conclude that compared with TS-RBHR, the energy utilization rate of EERPMS increased by 49.25% when the network energy is consumed $30J$. With reference to the network lifetime, Fig. 24 illustrates the FDN, HDN and LDN of EERPMS and TS-RBHR. Except LDN, the FDN and HDN of EERPMS outperform TS-RBHR. Based on the above experimental results, EERPMS can improve the utilization rate of network energy and effectively monitor the target area for a long time. The main reason is that the area division of TS-RBHR will cause the communication distance between CH and CMs to be too large (for example, in region 3, the maximum between CH and CM is 170 $m$ and far greater than the distance threshold $d_{th}$), which will lead to excessive energy consumption. In addition, $TH_1$, $TH_2$, $TH_3$ and $TH_4$ are not a variable threshold, which can't make the network dynamically select the appropriate CHs according to the changes of the residual energy and location of nodes.

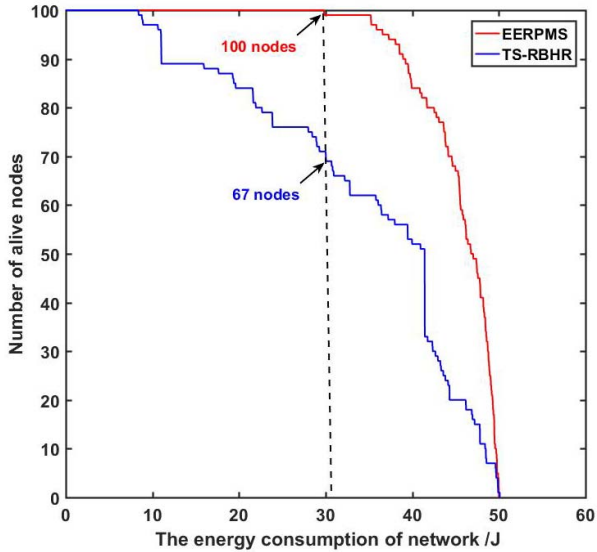

Fig. 23. The number of alive nodes varies with network energy consumption.

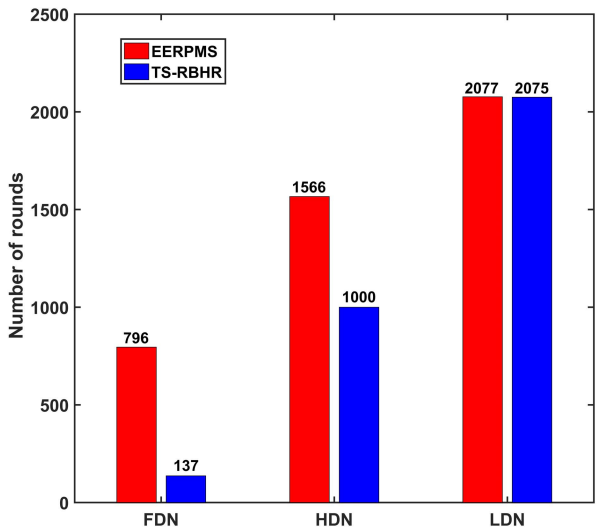

Fig. 24. The comparison of network lifetime.

## VI. CONCLUSION

In this paper, we present a clustering routing protocol based on multi-threshold segmentation. At first, a new cluster formation method was proposed to improve the distribution uniformity and load balancing of CHs. Then, we presented a theory to construct the relationship among CH location and number, and network energy consumption, which was verified and analyzed in Section IV. In addition, the CH selection algorithm based on the optimal location of CHs and the residual energy of nodes was adopted. Verified by simulation experiment, the proposed algorithm consumes less network energy and prolongs the network lifetime. It will provide an effective method for routing protocol design in IoT-based precision agriculture. Further, this work can be further exploited to support multi-hop routing in the case of monitoring very large areas and we infer that how to determine the optimal CH location in each sector area to minimize network energy consumption is a research challenge since multiple cluster heads should be selected in each cluster. Moreover, mobile sensors network [31] and BS with mobility are promising directions for our future work to provide services for more application scenarios.

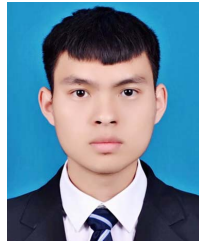

**Xiong Li** (Student Member, IEEE) is pursuing the master's degree with the Xi'an University of Posts and Telecommunications, Xi'an, China. His research interests include the technology and application of the Internet of Things.

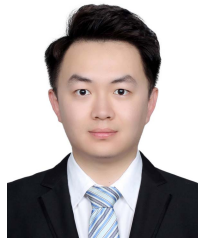

**Yan-Peng Cui** (Student Member, IEEE) received the M.S. degree from the Xi'an University of Posts and Telecommunications, Xi'an, China, in 2020. He is currently pursuing the Ph.D. degree with the School of Information and Communication Engineering, Beijing University of Posts and Telecommunications. His research interests include the technology and application of the Internet of Things and UAV networks.

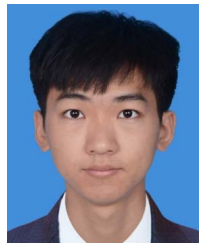

**Jia-Jun Wang** is pursuing the master's degree with the Xi'an University of Posts and Telecommunications, Xi'an, China. His research interests include the technology and application of the Internet of Things.

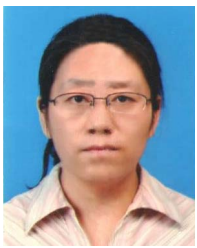

**Yin-Di Yao** was born in Baoji, China, in April 1978. She received the M.S. degree in communications and information systems. She is a Senior Engineer with the Xi'an University of Posts and Telecommunications, Xi'an, China. Her current research interests include the technology and application of the Internet of Things.

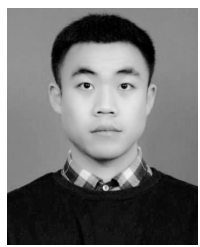

**Chen Wang** is pursuing the master's degree with the Xi'an University of Posts and Telecommunications, Xi'an, China. His research interests include the technology and application of the Internet of Things.